\documentclass{aa}
\usepackage{times,amssymb,amsmath}
\usepackage{epsfig}

\thesaurus{11 
           05.01.1 
	   08.19.5 
           13.07.1)} 

\title{Detection of the optical afterglow of GRB~000630: Implications
for dark bursts 
\thanks{Based on observations made with the Nordic Optical Telescope,
operated on the island of La Palma jointly by Denmark, Finland,
Iceland, Norway, and Sweden, in the Spanish Observatorio del 
Roque de los Muchachos of the Instituto de Astrofisica de 
Canaria, and with the German-Spanish Astronomical Centre, Calar Alto, 
operated by the MPI f\"{u}r Astronomie, Heidelberg, jointly with the 
Spanish National Commission for Astronomy, and with the TNG, operated 
on the island of La Palma by the Centro Galileo Galilei of the CNAA at the
Spanish Observatorio del Roque de los Muchachos of the Instituto de
Astrofisica de Canarias. 
}}

\author{
	J.U. Fynbo \inst{1,2}
        \and B.L. Jensen \inst{1}
	\and J. Gorosabel \inst{3}
	\and J. Hjorth \inst{1}
	\and H. Pedersen \inst{1}
	\and P. M{\o}ller \inst{2}
	\and T. Abbott \inst{4}
	\and A.J. Castro--Tirado\inst{5}
	\and D. Delgado \inst{6}
	\and J. Greiner \inst{7}
	\and A. Henden \inst{8}
	\and A. Magazz\`u \inst{9}
	\and N. Masetti \inst{10}
        \and S. Merlino \inst{5}
        \and J. Masegosa \inst{5}
	\and R. {\O}stensen \inst{11}
	\and E. Palazzi \inst{10}
	\and E. Pian \inst{10}
	\and H.E. Schwarz \inst{5}
	\and T. Cline \inst{12}
	\and C. Guidorzi \inst{13}
	\and J. Goldsten \inst{14}
	\and K. Hurley \inst{15}
	\and E. Mazets \inst{16}
	\and T. McClanahan \inst{12}
	\and E. Montanari \inst{13}
	\and R. Starr \inst{17}
	\and J. Trombka \inst{12}
}

\institute{
           Astronomical Observatory,
           University of Copenhagen,
           Juliane Maries Vej 30, DK--2100 Copenhagen \O, Denmark
	   \and
	   European Southern Observatory, 
	   Karl-Schwarsschild-Stra{\ss}e 2, D-85748 Garching,
	   Germany
	   \and
           Danish Space Research Institute,
           Juliane Maries Vej 30, DK--2100 Copenhagen \O, Denmark
	   \and
	   Nordic Optical Telescope,
	   Apartado Postal 474, E-38700 Santa Cruz de La Palma, Spain
	   \and
           Instituto de Astrof{\'\i}sica de Andaluc{\'\i}a, CSIC,
           Apartado Postal 3004, E-18080 Granada, Spain
           \and
	   Stockholm Observatory, SE--133 36 Saltsj{\"o}baden, Sweden
	   \and
           Astrophysikalisches Institut Potsdam, An der Sternwarte 16,
           14482 Potsdam, Germany
           \and
	   Universities Space Research Association,
	   U. S. Naval Observatory, Flagstaff Station, P. O. Box 1149,
	   Flagstaff, AZ 86002-1149, USA
	   \and
	   Telescopio Nazionale Galileo, Apartado Postal 565, E-38700 
	   Santa Cruz de La Palma, Spain
	   \and
	   Istituto Tecnologie e Studio Radiazioni Extraterrestri, 
	   CNR, Via Gobetti 101, 40129 Bologna, Italy
	   \and
	   Department of Physics, University of Troms{\o}, Troms{\o},  
	   Norway
           \and
           NASA Goddard Space Flight Center,
           Greenbelt, MD 20771, USA
	   \and
	   Universit{\'a} di Ferrara, Dipartimento di Fisica, Via Paradiso 12,
	   44100 Ferrara, Italy
	   \and
	   Applied Physics Laboratory, Johns Hopkins University, 
	   Laurel MD 20723, USA
           \and
           University of California, Berkeley,
           Space Sciences Laboratory,
           Berkeley, CA 94720--7450, USA
           \and
           Ioffe Physico-Technical Institute, St. Petersburg, 
           194021 Russia 
	   \and
	   Dept. of Physics, Catholic University of America,
	   Washington, DC 20064, USA
           }

\offprints{J.U. Fynbo}
\mail{jfynbo@eso.org}

\date{Received  / Accepted }

\begin{document}

\maketitle

\begin{abstract}
We present the discovery of the optical transient of the
long--duration gamma-ray burst GRB~000630. The optical 
transient was detected with the Nordic Optical Telescope
21.1 hours after the burst. At the time of discovery 
the magnitude of the transient was R = 23.04$\pm$0.08. The
transient displayed a power-law decline characterized by
a decay slope of $\alpha$ = $-$1.035$\pm$0.097. A deep image
obtained 25 days after the burst shows no indication of
a contribution from a supernova or a host galaxy
at the position of the transient. The closest 
detected galaxy is a R=24.68$\pm$0.15 galaxy 2.0
arcsec north of the transient.  
The magnitudes of the optical afterglows of GRB~980329, 
GRB~980613 and GRB~000630 were all R$\gtrsim$23 less than 24 
hours from the burst epoch. We discuss the implications of 
this for our understanding of GRBs without 
detected optical transients. We conclude that {\it i)}
based on the gamma-ray properties of
the current sample we cannot conclude that GRBs with no 
detected OTs belong to another class of GRBs than GRBs with 
detected OTs and {\it ii)} 
the majority ($\gtrsim$~75\%) of GRBs for which searches for 
optical afterglow have been unsuccessful are consistent with 
no detection if they were similar to bursts like GRB~000630 
at optical wavelengths. 
\end{abstract}

\section{Introduction}

The discoveries of the first $X$-ray afterglow (Costa et al. 1997) and 
Optical Transient (OT) (van Paradijs et al. 1997) of a gamma-ray burst 
(GRB) have led to a major breakthrough in GRB research. The determination of 
a redshift of 0.835 for \object{GRB~970508} (Metzger et al. 1997), and
the subsequent determination of redshifts of more than a dozen bursts 
with a median redshift of $\sim$1.0, have firmly established their 
cosmological origin (e.g. Kulkarni et al. 2000 and references therein).

In several cases supernovae (SNe) have been connected to GRBs. Attention to
this connection was, from the observational side, first drawn in the
remarkable case of \object{GRB~980425} that was associated with
\object{SN1998bw} (Galama et al. 1998). Furthermore, the light--curves 
of GRB~970228 and GRB~980326 (Reichart 1999, Galama et al. 2000; 
Castro-Tirado and Gorosabel 1999; Bloom et al. 1999) have bumps that are 
consistent with a contribution from an underlying SN. In no cases the 
presence of an underlying SN contribution to the light--curve has been 
firmly excluded. The exclusion of a SN contribution from
the light--curve of GRB~990712 (Hjorth et al. 2000) was based on a wrong 
identification of the OT in the {\it HST} image. A revised analysis
indeed shows an underlying SN contribution to the light--curve 
(Bj{\"o}rnsson et al. 2001).

Moreover, for less than half ($\approx$ 30\%) of all GRBs with 
well-determined coordinates have searches for optical counterparts been 
successful. In some cases the lack of detection can be easily explained by
too bright detection limits (e.g. due to light from the moon, nearby bright
stars or twilight), too slow reaction time or high foreground extinction
in the Galaxy. However, GRB~981226 (Frail et al. 1999) and GRB~990506 
(Taylor et al. 2000) displayed no optical afterglow down to R--band
limits of R=23 and R=23.5 less than half a day after the bursts (Jensen et
al. 1999; Pedersen et al. 1999), and both had a detectable radio afterglow. 
GRB~970828 (Groot et al. 1998a) had a weak X-ray counterpart, but no OT was 
detected down to an R-band limit of 23.8 only 3 hours after the burst. 
It is still not known whether these `dark' bursts constitute the faint
end of a (continuous) OT luminosity function or whether they represent a 
special class of GRBs that are truly dark in the optical (in other
words a dichotomy similar to the radio-loud vs. radio quiet dichotomy
for QSOs).

In this paper we present the detection of the OT of GRB~000630.
GRB~000630 was observed by Ulysses, Konus-Wind, NEAR, and the 
BeppoSAX GRBM June 30.02 UT (Hurley et al. 2000). 
GRB~000630 had a duration of
$\sim$20 s, and a 25--100 keV fluence of $\sim$2$\times$10$^{-6}$ erg 
cm$^{-2}$. These gamma-ray properties are consistent with the 
long--duration class of bursts (Hurley 1992; Kouveliotou et al. 1993). In 
Sect.~\ref{observations} and in Sect.~\ref{results} we present our 
optical observations and the results we have obtained. We also
discuss the 
implications of our results on our understanding of the $\approx$ 70\%
of GRBs with no detected optical counterparts. Finally we summarize 
our results in Sect.~\ref{summary}.

\section{Observations}
\label{observations}

 The IPN error-box of \object{GRB~000630} (Hurley et al. 2000) 
was observed in the R--band with the 2.56-m Nordic Optical Telescope 
(NOT) on 2000 June 30.90 UT (0.88 days after the burst) using the 
Andaluc\'{\i}a Faint Object Spectrograph (ALFOSC). Comparing with red 
and blue Palomar Optical Sky Survey~II exposures no  
OT was found (Jensen et al. 2000a). Further R--band imaging of the
field was carried out on July 1.9 UT. One object was found to have faded 
about one magnitude from June 30.90 to July 1.9. Subsequent deep R--band 
imaging obtained on July 3.9 UT confirmed the transient nature of the
object and hence established this object as the likely optical afterglow
of GRB~000630 (Jensen et al. 2000b). Further deep R--band imaging of 
the OT was obtained at the NOT on July 10.9 UT and on July 25.9 UT. 
We have also obtained R--band imaging with the U.S. Naval Observatory 
Flagstaff Station (USNOFS) 1.0-m  telescope and the Calar Alto (CA) 
2.2-m telescope (Greiner et al. 2000), 
and B and V images with the 3.5-m Telescopio Nazionale Galileo (TNG). 
The journal of observations is reported in Table~\ref{obslog}.

\begin{table}[t] 
\begin{small}
\begin{center}
\caption{The journal of observations.}
\begin{tabular}{@{}lllcccc}
UT         & Telescope & Band & Magnitude & Seeing    & Exp. time \\
           &           &        &           & (arcsec)  & (sec) \\
\hline
June 30.87 & CA          &  R    & 23.00$\pm$0.30 & 1.4  & 2$\times$600 \\
June 30.90  & NOT        &  R   & 23.04$\pm$0.08 & 0.9  & 3$\times$300 \\
July  1.24  & USNO       &  R & 23.13$\pm$0.25 & 2.1  & 18$\times$600 \\
July  1.88  & NOT        &  R   & 24.05$\pm$0.16 & 1.2  & 3$\times$600 \\
July  1.89  & CA         &  R   & 23.85$\pm$0.22 & 1.3  & 2$\times$900 \\
July  3.91  & NOT        &  R   & 24.67$\pm$0.14 & 0.8  & 5$\times$600 \\
July  4.06  & TNG        &  B   & $>$26.0        & 1.0  & 1800 \\
July  4.10  & TNG        &  V   & 25.44$\pm$0.23 & 1.3  & 1800 \\
July  10.9  & NOT        &  R   & $>$25.3        & 1.1  & 6$\times$600 \\
July  25.9  & NOT        &  R   & $>$26.1        & 0.8  & 11$\times$600 \\
\hline
\label{obslog}
\end{tabular}
\end{center}
\end{small}
\end{table}

\section{Results}
\label{results}

\subsection{Astrometry}
\begin{figure}
\begin{center}
\epsfig{file=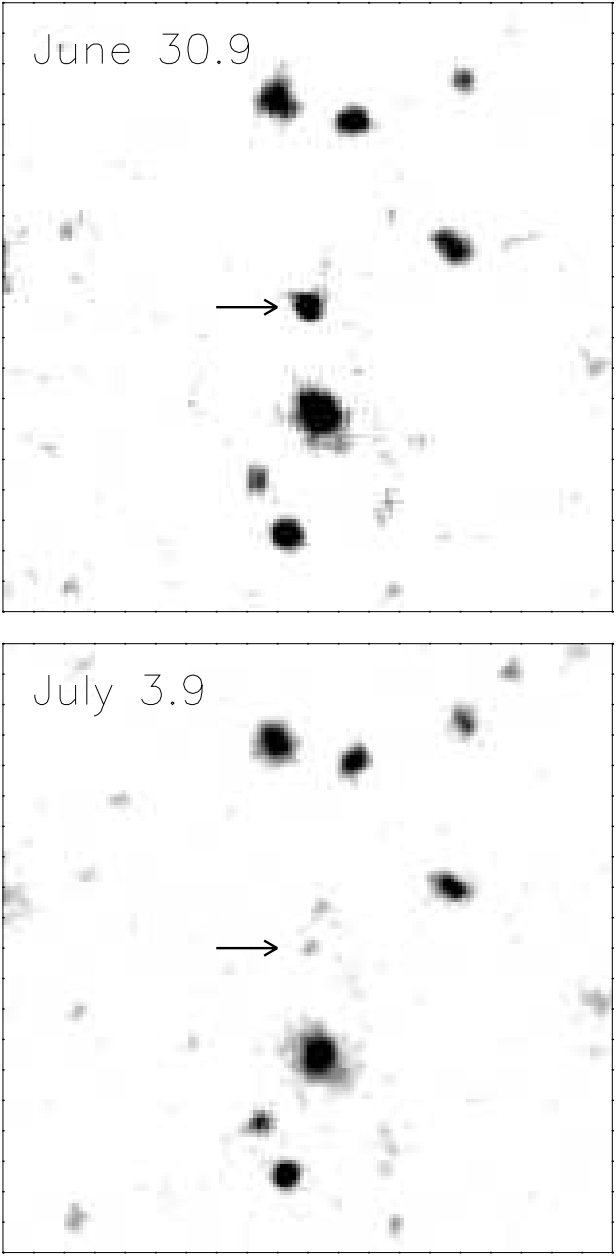,width=8.5cm}
\caption{{\it Upper panel\/}: A 30$\times$30 arcsec$^2$ section 
of the combined image from June 30.9, about 21.1 hours after the 
burst, centered on the Optical Transient (OT) of 
\object{GRB~000630} (marked by an arrow). {\it Lower panel\/}: 
The same region on July 3.9. The OT has faded by about 1.7 
magnitudes. Note also the faint galaxy 2 arcsec north of the OT.
}
\label{OT}
\end{center}
\end{figure}

By measuring the position of the OT relative to 29 stars in the 
USNO-A2.0 catalog we found the celestial coordinates of the OT to be 
RA(J2000) = 14:47:13.485, Dec(J2000) = +41:13:53.25. The rms
deviations around a fit to the positions of the 29 stars is 0.3 arcsec. 
A region of the NOT images centred on the OT is shown in Fig.~\ref{OT}.

\subsection{Photometry}
\subsubsection{The R--band light--curve}
As the OT was faint at the time of discovery and as it is 
situated in a region with several nearby galaxies we
have performed Point Spread Function (PSF) photometry with DAOPHOT~II
(Stetson~\cite{S1987},~\cite{S1997}) to measure the magnitude
of the OT. We first performed relative PSF-photometry between the 
OT and stars in the field. Standard magnitudes for these stars were
then obtained from Henden (2000) and the offset from the 
PSF-photometry to the standard system was derived. The
magnitudes of the OT derived in this way are given in 
Table~\ref{obslog}. The OT was not detected in the two latest epochs
(July 10.9 and 25.9). For these epochs we have provided the 2$\sigma$
upper limits to the OT R--band magnitude. 

To derive the light--curve parameters we fit a linear relation of 
the form $R = R_0 - \alpha  2.5 \log(\Delta t)$
to the NOT and Yost et al. (2000) data points. The $\chi^2$ per degree 
of freedom is 0.56 and the derived value of the power-law decay 
slope is $\alpha$ = $-$1.035$\pm$0.097. 

\subsubsection{B and V band photometry}
We performed the B and V--band photometry in the same way as for the
R--band. The results are shown in Table~\ref{obslog}. The OT was only 
detected in the V--band, hence for the B filter we can only derive a 
2$\sigma$ upper limit. The V$-$R colour of the OT is 0.77$\pm$0.27,
redder than most OTs observed so far and similar to the red afterglow
of GRB~000418 (Klose et al. 2000). The foreground extinction towards 
the field of GRB~000630 is E(B$-$V)=0.013$\pm$0.02 (Schlegel et al. 
1998), which is negligible.

\subsection{Limit on the magnitude of the host galaxy}
There are no indications of a host galaxy right underneath
the point source emission from the OT. The
closest galaxy is located 2.0 arcsec north of the OT. In order to
put a stringent limit on the magnitude of any galaxy
coincident with the OT we combined all the images
obtained on July 10.9 and July 25.9 (a total of 2.8 hours of 
imaging) using the code described in M{\o}ller and Warren (1993). 
The final combined image reaches a 5$\sigma$ limiting magnitude of 
25.2 in a 1.0 arcsec radius circular aperture. The FWHM of point
sources in the combined image is 0.93 arcsec.

\begin{figure}
\begin{center}
\epsfig{file=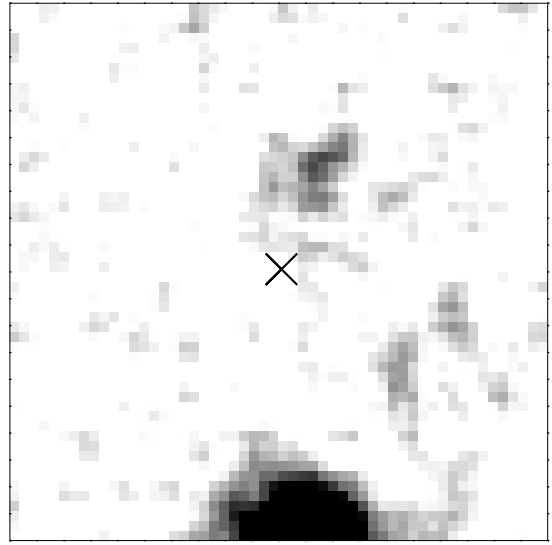,width=7.5cm}
\caption{A 10$\times$10 arcsec$^2$ section of the combined image 
from July 10.9 and July 25.9 centered on the position of the OT 
of \object{GRB~000630}. The OT is not seen (the position has been 
marked by $\times$). A galaxy is detected 2.0 arcsec north of the 
OT. The image has been smoothed with a 3$\times$3 pixel box-car
filter.
}
\label{host}
\end{center}
\end{figure}

Fig.~\ref{host} shows a 10$\times$10 arcsec$^2$ region centered on
the position of the OT. No emission is detected at that position.
The magnitude of the nearest galaxy, 2.0 arcsec north of the OT,
is 24.68$\pm$0.15 measured in a 1.0 arcsec radius circular aperture.
The galaxy has an extension in the direction of the OT, but we 
cannot with any certainty establish whether this galaxy is the host
galaxy of GRB~000630. The galaxy is also detected in the B and V  
band images of TNG. The colours of the galaxy are B$-$V=0.1$\pm$0.3 
and V$-$R=0.2$\pm$0.2, which is blue compared e.g. to the galaxies
in the NTT deep field (Arnouts et al. 1999, their Fig.~2).

\subsection{Constraint on the presence of an underlying SN}

The OT was not detected in the
deep R--band image obtained 25.9 days after the event, which 
constrains
the possible contribution to the light--curve from a supernova 
similar to SN1998bw. Even if the decay slope $\alpha$ has
steepened to $\alpha$ = $-$2.5 four days after the high energy
event we can exclude (by 2 and 0.5 magnitudes for redshifts
of z=0.5 and 1.3 respectively) a contribution from a supernova 
like SN1998bw if the redshift of the GRB is below $\sim$1.3. 
Any extinction in the host galaxy of the OT emission will 
of course loosen this constraint.

\subsection{Implications for dark bursts}

The OTs of GRB~980329, GRB~980613 and GRB~000630 
are the faintest optical transients detected within 24 hours from 
the high energy 
events. The OT of GRB~980329 was detected with the New Technology
Telescope at R=23.6$\pm$0.2 0.83 days after the GRB (Palazzi et al. 
1998) at the position of a previously found radio afterglow
(Taylor et al. 1998). The OT of GRB~980613 was detected with the NOT 
0.69 days after the GRB with a magnitude of R=22.9$\pm$0.2 (Hjorth et al. 
1998, 1999; Hjorth et al. in prep.). GRB~000630 is interesting 
because for this faint burst the afterglow properties (decay and 
spectral slope) are fairly well constrained.
The faintness of the OT of GRB~000630 cannot be explained by a steep 
decay slope as in the case of GRB~980326 (Groot et al. 1998b).
Hence, the OT must be faint either due to reddening in 
the host galaxy as in the case of GRB~000418 (Klose et al. 2000), 
a very high redshift as in the case of 
GRB~000131 (Andersen et al. 2000) or a low-density circumburst 
environment (M{\'e}sz{\'a}ros and Rees 1997; Taylor et al. 2000). 
The V$-$R colour of the OT implies a spectral slope of 
$\beta=-2.09\pm$0.97 (F$_{\nu}$ $\propto t^{\alpha} 
\nu^{\beta}$), which, though uncertain, may indicate either
significant reddening in the host galaxy or a high redshift.
Spectral slopes of GRBs are typically between $-$0.6
and $-$1.0.

\begin{table}[h]
\begin{center}
\begin{small}
\caption{A compilation of upper limits to magnitudes and observation 
times after the high energy event ($\Delta t$) for 43 GRBs with no 
detection of an OT. Besides IAU and GCN circulars the references
are: (1) Gorosabel et al. 1998; (2) Gorosabel 1999; (3) Groot et al.
1998a; (4) Gorosabel 2000.}
\begin{tabular}{@{}llllccc}
Burst & R-limit & $\Delta t$ & Reference \\
      & (mag)   & (days)     &           \\
\hline
\object{GRB~970111} &  23.0  &       0.79   &  (1) \\     
\object{GRB~970402} &  23.0  &       1.05   &  (2) \\     
\object{GRB~970616} &  20.0  &       2.64   &  IAUC 6687 \\     
\object{GRB~970815} &  23.0  &       0.70   &  IAUC 6721 \\     
\object{GRB~970828} &  23.8  &       0.14   &  (3) \\     
\object{GRB~971227} &  20.5  &       0.56   &  IAUC 6800 \\     
\object{GRB~981220} &  19.5  &       2.13   &  GCN 159, 165 \\     
\object{GRB~981226} &  23.0  &       0.4    &  GCN 190       \\           
\object{GRB~990217} &  22.5  &       0.28   &  GCN 258, 262 \\
                    &  23.5  &       0.83   &  GCN 262 \\
\object{GRB~990316} &  19.5  &       0.33   &  GCN 276, 277 \\
                    &  24.3  &       1.78   &  GCN 280 \\
\object{GRB~990506} &  23.5  &       0.49   &  GCN 291, 352  \\    
\object{GRB~990527} &  22.0  &       1.8    &  GCN 347, 349 \\
\object{GRB~990627} &  21.0  &       0.94   &  GCN 356, 358 \\
\object{GRB~990704} &  18.0  &       0.14   &  GCN 360, 362 \\
                    &  22.5  &       0.19   &  GCN 371 \\
\object{GRB~990806} &  23.5  &       0.52   &  GCN 392, 396 \\
\object{GRB~990907} &  23.0  &       1.64   &  GCN 405, 413 \\
\object{GRB~990908} &  20.0  &       0.47   &  GCN 406, 408 \\
\object{GRB~990915} &  20.5  &       1.25   &  GCN 410, 416 \\
\object{GRB~991014} &  23.1  &       0.47   &  GCN 417, 423 \\
\object{GRB~991105} &  23.5  &       0.65   &  GCN 433, 449 \\
\object{GRB~991106} &  22.0  &       0.42   &  GCN 435, 440 \\
\object{GRB~000115} &  22.0  &       1.52   &  GCN 519, 524 \\
\object{GRB~000126} &  19.0  &       2.5    &  GCN 525, 527 \\
\object{GRB~000301A}&  21.5  &       1.11   &  (4)          \\
\object{GRB~000307} &  22.0  &       2.11   &  GCN 601, 617 \\
\object{GRB~000315} &  19.0  &       1.29   &  (4)          \\
\object{GRB~000323} &  21.5  &       2.04   &  GCN 616, 621 \\
\object{GRB~000326} &  21.5  &       3.19   &  GCN 618, 625 \\
\object{GRB~000408} &  21.5  &       1.0    &  GCN 626, 633 \\
\object{GRB~000424} &  20.7  &       1.36   &  GCN 644, 648 \\
                    &  22.8  &       2.44   &  GCN 660 \\
\object{GRB~000429} &  19.0  &       0.73   &  GCN 657, 659 \\
\object{GRB~000508B}&  22.5  &       1.4    &  GCN 665, 670 \\
\object{GRB~000519} &  21.0  &       0.82   &  GCN 672, 679 \\
\object{GRB~000528} &  22.3  &       0.53   &  GCN 675, 674 \\
                    &  23.3  &       0.73   &  GCN 691 \\
\object{GRB~000529} &  19.8  &       1.10   &  GCN 676, 682 \\
\object{GRB~000604} &  22.0  &       2.24   &  GCN 687, 692 \\
                    &  23.0  &       1.90   &  (4)          \\
\object{GRB~000608} &  20.5  &       0.59   &  (4)          \\
\object{GRB~000615} &  21.5  &       0.18   &  GCN 703, 709 \\
\object{GRB~000616} &  18.0  &       1.6    &  GCN 711, 714 \\
\object{GRB~000620} &  19.8  &       0.24   &  GCN 722, 734 \\ 
                    &  21.7  &       0.67   &  GCN 734 \\
\object{GRB~000623} &  20.0  &       1.21   &  GCN 730, 735 \\ 
\hline
\label{darkdata}
\end{tabular}
\end{small}
\end{center}
\end{table}

\begin{figure}
\begin{center}
\epsfig{file=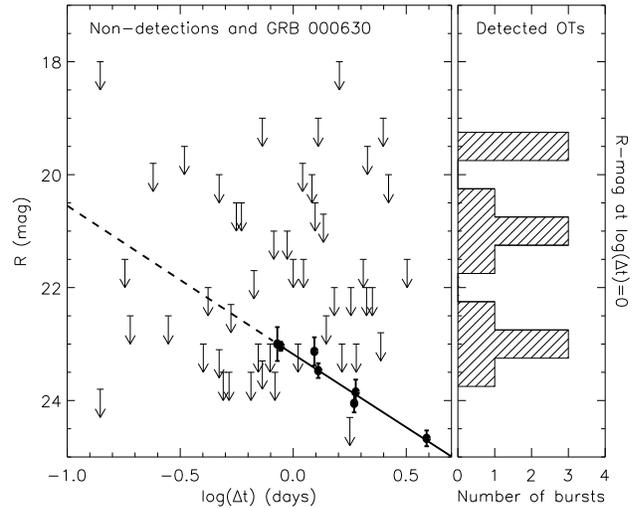,width=8.5cm}
\caption{{\it Left panel:} the limiting R--band magnitudes vs. the 
logarithm of the number of days from the GRB epoch to the time of 
observation for 43 GRBs with no detected OT. Also plotted is the 
light--curve of the OT of GRB~000630. The extrapolation to earlier 
times than we have data for is shown as a dashed line.
{\it Right panel:} a histogram of R--band magnitudes at 
$\Delta t$ = 1 day for GRBs with detected OTs (GRBs prior to 
and including GRB~000630).
}
\label{dark}
\end{center}
\end{figure}

It is still not understood why OTs have only been detected for about
30\% of well localized GRBs. Possible explanations are {\it i)} there
is a optically bright vs. optically dark dichotomy for GRBs similar
to the radio quiet vs. radio loud dichotomy for QSOs (i.e. a bimodal
OT luminosity function), {\it ii)} a large fraction of GRBs 
occur at redshifts $z \gtrsim$ 7 and are hence invisible in the 
optical due to Ly-$\alpha$ blanketing and absorption by intervening 
Lyman-limit systems, {\it iii)} a large fraction of GRBs occur in highly
obscured galaxies similar to the SCUBA-selected galaxies (e.g. Ivison
et al. 2000) so that the optical emission never escapes the host 
galaxies, or {\it iv)} the shape of the OT luminosity function is 
such that with the search strategies applied in the period 1997--2000 we
would not expect to detect OTs for more than 30\% of the well localized
GRBs.

In order to shed some light on this problem we have compiled in 
Table~\ref{darkdata}, mainly using the GCN Circular 
archive\footnote{{\tt {\tiny
http://gcn.gsfc.nasa.gov/gcn/gcn3\_archive.html}}}, the most stringent 
R--band upper limits to magnitudes and their observation epochs relative 
to the GRB epoch ($\Delta t$) for 43 GRBs with no detected OT.
The left panel in Fig.~\ref{dark} shows the upper limits plotted 
against $\Delta t$
for the 43 GRBs. Also plotted is the light--curve of the OT of
GRB~000630 (full-drawn line) and its extrapolation to earlier
times than the first observing epoch (dashed line). As seen, 
except for GRB~990316, all upper limits are above the 
{\it observed} ($\gtrsim$ 1 day after GRB event) light--curve of 
GRB~000630. However, the strong 
upper limit on the magnitude of GRB~990316 is only valid for a  
2$\times$2 arcmin$^2$ region around a candidate LOTIS transient
that may be unrelated to the GRB. The upper limits for GRB~970815, 
GRB~970828, GRB~981226, GRB~990217, GRB~990506, GRB~990704, GRB~990806, 
GRB~991014, GRB~991105, and GRB~000615 are below the 
extrapolation of the light--curve to earlier times. Two of these,
GRB~991014 and GRB~991105, occurred in fields with relatively
high foreground extinction (with E(B$-$V) values 0.268 and 0.598 
respectively). Note here, that not all OT light--curves
peak at $\Delta$t$<$1 day (Guarnieri et al. 1997, but see also
Galama et al. 2000; Pedersen et al. 1998). In the right panel
of Fig.~\ref{dark} we show the histogram of R--band magnitudes
at $\Delta$t=1 day (in 0.5 mag bins) of the 13 OTs (prior to and
including GRB~000630) that were 
detected earlier than 24 hr after the burst. It is much more difficult 
to detect an OT at R=23 than at R=19.5, and an OT having R=24 at 
$\Delta$t=1 day would almost certainly not be detected with current 
search strategies. Therefore the observed distribution is {\it not} 
the real distribution of apparent OT magnitudes. The real distribution 
must have a larger fraction of bursts in the fainter bins. From the two 
plots in Fig.~\ref{dark}
we conclude that the majority ($\gtrsim$ 75\%) of GRBs without 
detected OTs are consistent with no detection if they were similar 
to dim bursts like GRB~980329, GRB~980613, and GRB~000630 at 
optical wavelengths. 

\begin{figure}
\begin{center}
\epsfig{file=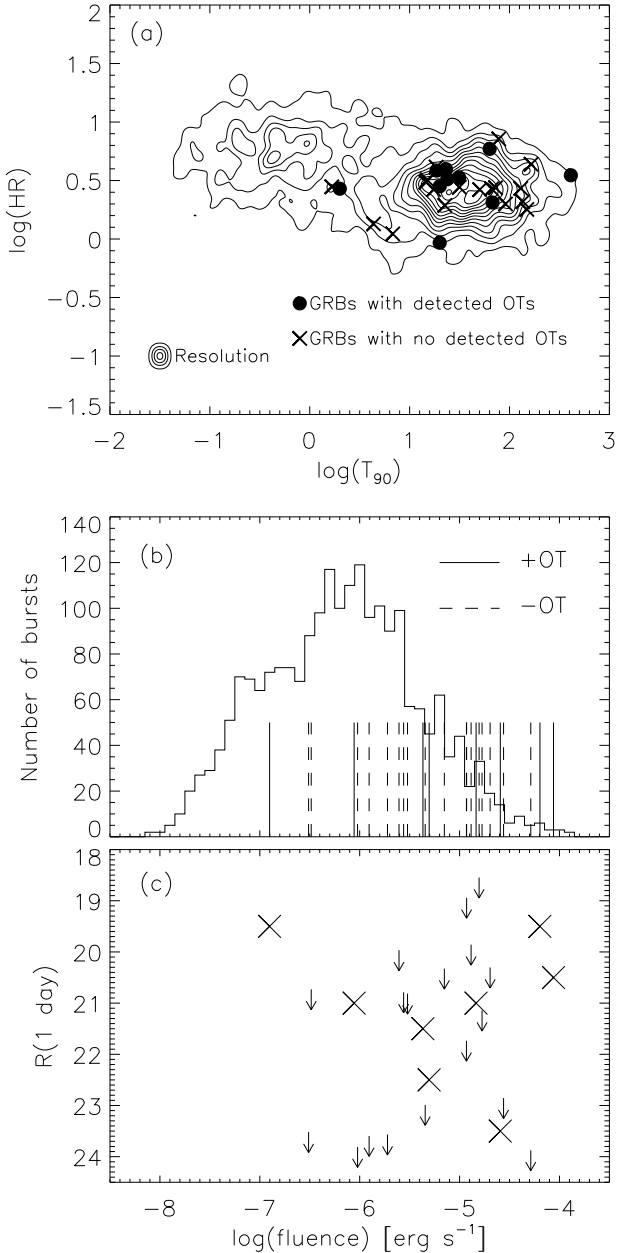,width=8.5cm}
\caption{{\it a}: The spectral hardness vs. duration plot for 1959 BATSE
bursts. Indicated are the positions in the diagram of the GRBs with
and without detected OTs.
{\it b}: The channel 3 fluence histogram for all BATSE bursts and the
positions in the distribution of GRBs with and without detected OTs.
{\it c}: The R(1 day) plotted against the log of the channel 3 fluence. 
}
\label{compare}
\end{center}
\end{figure}

In Fig.~\ref{compare}a--c we compare the properties of BATSE-detected GRBs 
with and without detected OTs. In Fig.~\ref{compare}a we show the 
spectral hardness vs. duration plot for 1959 BATSE bursts (shown as
contours following Jensen et al. (2001)). Superimposed as filled circles 
and crosses are the positions
of GRBs with and without detected OTs. As seen, the two groups populate
the same region in the long-duration part of the diagram. In
Fig.~\ref{compare}b we show the channel 3 fluence histogram in 0.1 dex bins
for all BATSE bursts. The GRBs with and without detected OTs again show
similar distributions. Finally, in Fig.~\ref{compare}c the R--band magnitude 
of the OT at $\Delta t$ = 1 day in 0.5 mag bins for GRBs with detected
OTs or the upper limit for GRBs without detected OTs are plotted against
the logarithm of the channel 3 fluence. We have shifted the upper limits
to $\Delta t$ = 1 day assuming a decay slope of $\alpha = -1$. Intuitively 
one might expect a
correlation between the energy released at gamma-ray wavelengths and in
the optical {\it at some level}, although different redshifts and different
amounts of dust obscuration will increase the scatter. However,
Fig.~\ref{compare}c does not indicate any correlation. Moreover, the
upper limits for the majority of GRBs with no detected OTs do not place them 
in regions of the diagram that are not already populated by GRBs with
detected OTs. This means that based on the gamma-ray properties of
the current sample we cannot 
conclude that GRBs with no detected OTs belong to another class of
GRBs than GRBs with detected OTs.

\section{Summary}
\label{summary}

The OT of GRB~000630 was detected with the NOT 21.1 
hours after the high energy event with an R--band magnitude of 
23.04$\pm$0.08. The OT subsequently
followed a power-law decay characterized by a slope 
of $\alpha$ = $-$1.035$\pm$0.097. 
From the deep R--band image obtained 25.9 days after the event we
find that there is no evidence for an underlying SN. Furthermore,
there is no indication of a host galaxy down to a 5$\sigma$ limit of
R=25.2 right underneath the point source emission from the OT. 
The closest galaxy is a blue galaxy located 2.0 arcsec north of the
OT.

For the problem of dark bursts we find that the majority
($\gtrsim$~75\%) of GRBs for which searches for
optical afterglow have been unsuccessful are consistent with
no detection if they were similar to bursts like GRB~000630
at optical wavelengths. We at present cannot exclude nor
confirm any of the explanations {\it i -- iv} listed above
(see Sect~3.5) for why some bursts are darker than others. In 
order to establish the reasons why most
GRBs searches for OTs up till now have been unsuccessful, and thereby obtain
a more complete understanding of optical afterglows to GRBs and the
relation between optical and high energy properties of GRBs, it is therefore 
essential to conduct deep (R$_{lim} \gtrsim$ 24 at $\Delta t\lesssim$1.0 
day) follow-up imaging 
at optical wavelengths and also deep infra-red follow-up of a sample 
of well localized and homogeneously selected GRBs. This is feasible with 
the next generation of GRB satellites and 8--m class telescopes.

\section*{Acknowledgments}
We thank our anonymous referee for comments that helped us
improve the discussion of dark bursts significantly.
Some of the data presented here have been taken using ALFOSC, which is
owned by the Instituto de Astrofisica de Andalucia (IAA) and operated 
at the Nordic
Optical Telescope under agreement between IAA and the NBIfAFG of the
Astronomical Observatory of Copenhagen.
Support for this programme by the director of the Nordic Optical
Telescope, professor Piirola, is much appreciated. We acknowledge
M. Feroci for his assistance with IPN operations and F. Frontera
for his assistance with BeppoSAX operations. 
KH is grateful for Ulysses support under JPL Contract 958056, for 
NEAR support under NASA grant NAG5-9503, and for BeppoSAX support 
under NASA grant NAG5-9126. We thank F. Hoyo for help in the 
observations at Calar Alto. JG acknowledges the receipt of a Marie 
Curie Research Grant from the European Commission. This work was 
supported by the Danish Natural Science Research Council (SNF).


\begin{thebibliography}{99}
\bibitem{AHP2000}Andersen M.I., Hjorth J., Pedersen H., et al., 2000, A\&A
364, L54
\bibitem{ADS2000} Arnouts S., D'Odorico S., Cristiani S., et al.,
1999, A\&A 343, 19.
\bibitem{BHJ2001}Bj{\"o}rnson G., Hjorth J., Jakobsson P., Christensen
L., Holland S., ApJL submitted
\bibitem{BKD1999}Bloom J.S., Kulkarni S.R., Djorgovski S.G., et al., 1999,
    Nature 401, 453
\bibitem{CTG1999}Castro-Tirado, A. J., \& Gorosabel, J., 1999, A\&AS, 138, 449
\bibitem{CFH1997}Costa E., Frontera F., Heise J., et al., 1997, Nature 387, 783
\bibitem{FKB1999}Frail D.A., Kulkarni S.R., Bloom J.S., et al., 1999, 
ApJ 525, L81
\bibitem{GAL}Galama T.S., Vreeswijk P.M., Paradijs J. van, et al. 1998, Nature
395, 670
\bibitem{GAL2}Galama T.S., Tanvir N., Vreeswijk P.M., et al. 2000, ApJ 536, 185
\bibitem{G1999}Gorosabel J., 1999, Ph D. Thesis, Univ. of Valencia.
\bibitem{GCW1998}Gorosabel J., Castro-Tirado A.J., Wolf C., et al., 1998,
A\&A 339, 719
\bibitem{G2000}Gorosabel J., 2000, Priv. Comm.
\bibitem{GHM2000}Greiner J., Henden A., Merlino S, et al., 2000, GCN 743
\bibitem{GGP1998}Groot P.J., Galama T.J., van Paradijs J., et al., 1998a, 
ApJ 493, L27
\bibitem{GGV1998}Groot P.J., Galama T.J., Vreeswijk P.M., et al., 1998b, 
ApJ 502, L123
\bibitem{GBM1997}Guarnieri A., Bartolini C., Masetti N., et al., 1997, A\&A 328, L13
\bibitem{H2000}Henden A., 2000, GCN 742
\bibitem{HAP1998}Hjorth J., Andersen M.I., Pedersen H., et al., 1998, GCN 109
\bibitem{HPJ1999}Hjorth J., Pedersen H., Jaunsen A.O., Andersen M.I., 1999,
A\&AS 138, 461
\bibitem{HHC2000}Hjorth J., Holland S., Courbin F., 2000, ApJL 534, 147
\bibitem{H1992}Hurley K., 1992, AIP Conf. Proc. 265, ``Gamma-Ray Bursts'', 
Eds. W. Paciesas and G. Fishman, AIP (New York)
\bibitem{HCM2000}Hurley K., Cline T., Mazets E., et al., 2000, GCN 736
\bibitem{ISB2000}Ivison R.J., Smail I., Barger A.J., et al., 2000, MNRAS 315, 209
\bibitem{JPH1999}Jensen B.L., Hjorth J., Pedersen H., et al., 1999, GCN 190
\bibitem{JPH2000}Jensen B.L., Pedersen H., Hjorth J., et al., 2000a, GCN 739
\bibitem{JFP2000}Jensen B.L., Fynbo J.P.U., Pedersen H., al., 2000b, GCN 747
\bibitem{JFG2001}Jensen B.L., Fynbo J.P.U., Gorosabel J., al., 2001, 
submitted to A\&A
\bibitem{KSM2000}Klose S., Stecklum B., Masetti N., et al., 2000, ApJ in
press
\bibitem{KMF1993}Kouveliotou C., Meegan C.A., Fishman G.J., et~al., 1993, ApJ 413, L101
\bibitem[2000a]{KBB2000}
Kulkarni S.R., Berger E., Bloom J.S., et al., 2000a, To appear in Proc.\ of the
5th Huntsville GRB Symposium (astro-ph/0002168)
\bibitem[1999]{MR1999} M\'esz\'aros P., Rees M.J., 1997, ApJ 476, 232
\bibitem[1997]{MDK1997}
Metzger M.R., Djorgovski S.G., Kulkarni S.R., et~al., 1997, Nature 387, 878
\bibitem[1993]{MW1993} M\o ller P., Warren S.J., 1993, A\&A 270, 43
\bibitem{PPM1998}Palazzi R., Pian E., Masetti N., 1998, A\&AL 336, 95 
\bibitem[1997]{PGG1997}
van Paradijs J., Groot P.J., Galama T., et~al., 1997, Nature 386, 686
\bibitem{PJG1998}Pedersen H., Jaunsen A.O., Grav T., et al., 1998, ApJ 496, 311
\bibitem{PHJ1999}Pedersen H., Hjorth J., Jensen B.L., Jaunsen A.O., Holland S.,
1999, GCN 352
\bibitem{Ric}Reichart D.,E., 1999, ApJ 521, L111
\bibitem[1998]{SFD1998}Schlegel D.J., Finkbeiner D.P., Davis M., 1998, 
ApJ 500, 525
\bibitem[1987]{S1987} Stetson P., 1987, PASP 99, 191S
\bibitem[1997]{S1997} Stetson P., 1997, ``User's Manual for DAOPHOT II''
\bibitem{TFK1998}Taylor G.B., Frail D.A., Kulkarni S.R., et al., 1998, 
ApJL 502, 115 
\bibitem{TBM2000}Taylor G.B., Bloom J.S., Frail D.A., et al., 2000, 
ApJ 537, L17 
\bibitem{YHD2000}Yost S., Harrison F., Diercks A., 2000, GCN 748
\end{thebibliography}
\end{document}